\documentclass[12pt]{article}
\usepackage{amsmath, amsfonts, epsfig, txfonts}
\def\theequation{\arabic{equation}}

   \def\theequation{\thesection.\arabic{equation}}
   \def\theequation{\arabic{equation}}

\newsavebox{\savepar}

\newtheorem{remark}{Remark}

\newtheorem{example}{Example}
\newtheorem{definition}{Definition}
\newtheorem{theorem}{Theorem}

\newtheorem{propos}{Proposition}

\def\appendix{\par
\setcounter{section}{0}
   \setcounter{equation}{0}
    \def\@chapapp{APPENDIX}
    \def\thesection{\Alph{section}}
    \def\theequation{A.\arabic{equation}}
    \def\section{
    \refstepcounter{section}
    \@startsection {section}{A}{0pt}{-3.5ex \@plus -1ex \@minus -.2ex}
                   {0.3ex \@plus.2ex}{\normalsize\sffamily\bfseries} }}

\def\eop{\hbox{{\vrule height7pt width3pt depth0pt}}}

\makeatletter
\newcommand{\least}{\let\CS=\@currsize\renewcommand{\baselinestretch}{1}\tiny\CS}

\newcommand{\oneandahalfspacing}{\let\CS=\@currsize\renewcommand{\baselinestretch}{1.2}\tiny\CS}
\newcommand{\doublespacing}{\let\CS=\@currsize\renewcommand{\baselinestretch}{2.5}\tiny\CS}

\usepackage{pst-plot}
\usepackage{amsmath}
\usepackage{amsfonts}
\usepackage{euscript}
\usepackage{oldgerm}

\usepackage{rotating}

   \renewcommand{\baselinestretch}{1.3}

\makeindex

\begin{document}

\newcommand{\namelistlabel}[1]{\mbox{#1}\hfil}
\newenvironment{namelist}[1]{%
\begin{list}{}
{
\let\makelabel\namelistlabel
\settowidth{\labelwidth}{#1}
\setlength{\leftmargin}{1.1\labelwidth}
}
}{%
\end{list}}

%\numberwithin{equation}{section}
\newcommand{\be}{\begin{equation}}
\newcommand{\ee}{\end{equation}}
\newcommand{\dist}{{\rm\,dist}}
\newcommand{\sspan}{{\rm\,span}}
\newcommand{\re}{{\rm Re\,}}
\newcommand{\im}{{\rm Im\,}}
\newcommand{\sgn}{{\rm sgn\,}}
\newcommand{\beano}{\begin{eqnarray*}}
\newcommand{\eeano}{\end{eqnarray*}}
\newcommand{\bea}{\begin{eqnarray}}
\newcommand{\eea}{\end{eqnarray}}

\newcommand{\ba}{\begin{array}}
\newcommand{\ea}{\end{array}}
\newcommand{\hone}{\mbox{\hspace{1em}}}
\newcommand{\hon}{\mbox{\hspace{1em}}}
\newcommand{\htwo}{\mbox{\hspace{2em}}}
\newcommand{\hthree}{\mbox{\hspace{3em}}}
\newcommand{\hfour}{\mbox{\hspace{4em}}}
\newcommand{\von}{\vskip 1ex}
\newcommand{\vone}{\vskip 2ex}
\newcommand{\vtwo}{\vskip 4ex}
\newcommand{\vthree}{\vskip 6ex}
\newcommand{\vfour}{\vspace*{8ex}}
\newcommand{\norm}{\|\;\;\|}
\newcommand{\integ}[4]{\int_{#1}^{#2}\,{#3}\,d{#4}}
\newcommand{\inp}[2]{\langle {#1} ,\,{#2} \rangle}
\newcommand{\vspan}[1]{{{\rm\,span}\{ #1 \}}}
\newcommand{\R} {{\mathbb{R}}}

\newcommand{\B} {{\mathbb{B}}}
\newcommand{\C} {{\mathbb{C}}}
\newcommand{\N} {{\mathbb{N}}}
\newcommand{\Q} {{\mathbb{Q}}}
\newcommand{\LL} {{\mathbb{L}}}
\newcommand{\Z} {{\mathbb{Z}}}

\newcommand{\BB} {{\mathcal{B}}}
\newcommand{\dm}[1]{ {\displaystyle{#1} } }
\def \stackt{{\stackrel{.}{.\;.}\;\;}}
\def \stackb{{\stackrel{.\;.}{.}\;\;}}
\def \olu{\overline{u}}
\def \olv{\overline{v}}
\def \olx{\overline{x}}
\def \olp{\overline{\partial}}
\def\diag{{\;{\rm diag } \; }}
\thispagestyle{empty}
\bibliographystyle{}

\newcommand{\alert}[1]{\fbox{#1}}

\begin{center}
{\Large \bf Average Collapsibility of Distribution Dependence and Quantile Regression Coefficients}\\
\end{center}

\vone \vtwo
%\begin{center}
\noindent {P. VELLAISAMY}\\
{\it Department of Mathematics, Indian Institute of Technology
Bombay}

\vone
\noindent {\it running headline: Average collapsibility}

\vone

\noindent {\bf ABSTRACT. The Yule-Simpson paradox notes that an
association between random variables can be reversed when averaged
over a background variable.  Cox and Wermuth  introduced the concept of distribution
dependence between two random variables $X$ and $Y$, and gave two
dependence conditions, each of which guarantees that reversal of
qualitatively similar conditional dependences cannot occur after
marginalizing over the background variable.  Ma, Xie and Geng
 studied the
uniform collapsibility of distribution dependence over a
background variable $W$, under stronger homogeneity condition.
Collapsibility ensures that associations are the same  for
conditional and marginal models. In this paper, we use the notion
of average collapsibility which requires only the conditional
effects average over the background variable to the corresponding
marginal effect and investigate  its conditions for distribution
dependence and for quantile regression coefficients.}

\vone {\noindent \it Key words:} {Average collapsibility,
collapsibility, conditional independence, contingency table,
distribution dependence, quantile regression coefficient,
Yule-Simpson paradox}.

\vone

\noindent {\bf 1. Introduction}

\vone \noindent There are several ways to interpret the association between a
response and an explanatory variable. The
association
may be measured by  odds ratio, or relative risk, or interaction
parameters of the corresponding log-linear model for categorical
variables, regression coefficient or distribution dependence for
continuous variables. The concept of collapsibility with respect
to these parameters was well studied by  Bishop (1971), Cox
(2003), Cox \& Wermuth (2003), Geng (1992), Ma \emph {et
al.~}(2006),
 Vellaisamy \& Vijay (2007, 2008, 2010), Wermuth (1987, 1989), Whittemore (1978) and Xie \emph {et al.~}
 (2008), among others. Cox \& Wermuth (2003) defined distribution
dependence as a measure of association between two variables,
 and discussed the
 effect reversal phenomenon, when a background variable (sometimes unobserved) is condensed. They obtained
 sufficient conditions
for no  effect reversal, that is, for the non-occurrence of Yule
\& Simpson's paradox. Recently, Ma \emph {et al.~}(2006) proved
that the conditions of Cox \& Wermuth (2003) are indeed necessary
and sufficient for uniform collapsibility of distribution
dependence, under the assumption that distribution dependence is
homogeneous over the background variable. \\

The concept of  average collapsibility for random coefficient models was
introduced and discussed in Vellaisamy \& Vijay (2008). In the
same spirit, this paper considers average collapsibility (A-collapsibility, henceforth) of distribution
dependence and quantile regression coefficients. Note that
A-collapsibility  means that the conditional effect averages over
the background variable to the corresponding marginal effect. The
conditions of Cox \& Wermuth (2003) are shown to be sufficient
for A-collapsibility, and also necessary when $W$ is a binary
variable. A necessary condition for A-collapsibility in terms of
conditional densities is also obtained. Recently, Cox (2007)
extended Cochran's result on regression coefficients of
conditional and marginal models to quantile regression
coefficients. The conditions of Cox \& Wermuth are also shown to
be sufficient for the A-collapsibility of quantile regression
coefficients.  We identify a class of conditional distributions of
$W$, given $Y=y$ and $X=x$, for which they are even necessary.
Applications to the analysis of a contingency table and linear
regression models are also considered.

\vone \noindent  {\bf 2. Collapsibility of distribution dependence}

\vone Let $X$ and $Y$ be two random variables. The dependence of $Y$ on
$X$ is called stochastically increasing if $P(Y >y\mid X= x)$ is
increasing in $x$ for all $y.$ That is,  when $X$ is continuous,
the partial derivative of  the conditional distribution function
$F(y\mid x)$  satisfies (Cox \& Wermuth, 2003)
\begin{eqnarray}
\frac{\partial{F(y\mid x)}}{\partial x} \leq 0,
\label{eq2}
\end{eqnarray}
for all $y$ and $x$, with strict inequality in a region of
positive probability.  Suppose also that $Y$ given $X=x$ and $W=w$
is stochastically increasing in $x$ for all $w,$ so that
 $ \displaystyle \frac{\partial{F(y\mid x, w)}}{\partial x} \leq 0$
for all $y$, $x$ and $w$. Then,
\begin{equation}
F(y\mid x)=P(Y \leq y\mid X=x)
                =  \displaystyle\int F(y\mid x, w) f(w\mid x)\,dw. \nonumber
\end{equation}
On differentiating with respect to $x$, we have
\begin{eqnarray} \label{eq2.2}
\frac{\partial{F(y\mid x)}}{\partial x}&=& \displaystyle\int \frac{\partial{F(y\mid x, w)}}{\partial x}
 f(w\mid x) \,dw
 + \int F(y\mid x,w) \frac{\partial{f(w\mid x)}}{\partial x} \,dw.
\label{eq2.2}
\end{eqnarray}

\noindent If $X\perp W$, then $f(w\mid x)= f(w)$ and so (Cox,
2003)
\begin{center}
$\displaystyle \frac{\partial{f(w\mid x)}}{\partial x}= 0,$
\end{center}
leading to
\begin{equation}
\label{neq2.3} \displaystyle\frac{\partial{F(y\mid x)}}{\partial
x}= \displaystyle\int \frac{\partial{F(y\mid x, w)}} {\partial x}
f(w)dw.
\end{equation}
When $X \perp W$, we have from (\ref{neq2.3}),
 $$\displaystyle\frac{\partial{F(y\mid x, w)}}{\partial x}\leq 0
\Longrightarrow \displaystyle\frac{\partial{F(y\mid x)}}{\partial x}\leq 0, \ {\rm{ for ~ all}}\
 y,x \ {\rm and} \ w. $$
 Thus, $Y$ remains stochastically increasing in $x$ after
marginalization over the covariate $W.$ Note in general
   (see (\ref{eq2.2})) it is possible that
$\displaystyle \frac{\partial{F(y\mid x, w)}}{\partial x}\leq 0$,
for all $y, x$ and $w$, but
 $\displaystyle \frac{\partial{F(y\mid x)}}{\partial x}>0$ for some $y$ and $x$, implying   effect reversal.
That is, the dependence of $Y$ and $X$ is no longer stochastically
increasing. This  effect reversal is known as Yule-Simpson paradox
(Cox \& Wermuth, 2003).

 \vone \noindent Let $Y$ be a
response variable, $X$ be an explanatory variable and $W$ be a
background variable. The function $ \displaystyle \frac{\partial
F(y|x,w)}{\partial x}$ is called a distribution dependence
function. When $X$ is discrete, the partial differentiation is
replaced by differencing between adjacent levels of $X$.  For
example, when $X$ is  ordinal with support $S(X)=\{1,\cdots,I\}$,
the distribution dependence function is defined as (Cox,
2003)
\begin{eqnarray}
\label{distcat}
\dfrac{\partial F(y|x,w)}{\partial x}=\Delta_x F(y|x,w)= P(Y\leq y\mid x+1,w) - P(Y\leq y\mid x,w),
\end{eqnarray}
for $x=1,2,\cdots,I-1$.
 The following definitions are due to Ma \emph{et al.~}(2006).

{\definition The distribution dependence function is said to be
homogeneous with respect to $W$ if
$$\dfrac{\partial F(y|x,w)}{\partial x}=\dfrac{\partial F(y|x,w')}{\partial x},$$
for all $y$, $x$ and $w\neq w'$.}

{\definition The distribution dependence function is said to be
collapsible over $W$ if
\begin{eqnarray}
\frac{\partial F(y|x, w)}{\partial x}=\frac{\partial F(y|x)}{\partial x},  \ {\rm{ for\  all}}\
 y,x \ {\rm and} \ w , \nonumber
\end{eqnarray}
and uniformly collapsible if
$$\frac{\partial F(y|x, W \in A)}{\partial x}=\frac{\partial F(y|x)}{\partial x}$$
for all $y, x $ and $A$ in the support of $W.$  When $W$ is
ordinal, the set $A$ is of the form $(i, i+1, \cdots, i+j).$} \\

Note that uniform collapsibility implies collapsibility, and
collapsibility implies homogeneity.  Homogeneity is commonly
assumed for pooled estimation as in Mantel \& Haenszel (1959).
 Ma \emph{et al.~}(2006)
showed that the distribution
 dependence function is uniformly collapsible iff either: (i) $Y \perp X | W$; or (ii) $X \perp W$ and $\displaystyle
\frac{\partial F(y|x,w)}{\partial x}$ is homogeneous in $w$. Cox
\& Wermuth (2003) noted that either condition (i) or (ii) is
sufficient to ensure that no  effect reversal can occur when
marginalizing the background variable $W$.

\newpage

\noindent {\bf 3. Average collapsibility of distribution dependence}

\vone \noindent A-collapsibility is a weaker condition  for non-reversal than
collapsibility. It requires only that the conditional effect
averages over the background variable to the corresponding
marginal effect, and does not require homogeneity. For example,
for a non-linear regression given $W$, the condition of
homogeneity over $W$ is not satisfied.

As a motivating example, we use the following  2x2x2 contingency
table where neither the homogeneity nor the collapsibility holds.

\begin{example} \label{ex-ct} {\em Consider the following $2\times 2\times 2$ table.
\[\begin{tabular}{|cc|lcr|}\hline
     &      &       &   $W$   &   \\
$X$  &   $Y$   &   1   &       &   2   \\\hline
    &   1   &   5  &       &   7  \\
1   &       &       &       &       \\
    &   2   &   7  &       &   3  \\\hline
    &   1   &   15  &       &   12  \\
2   &       &       &       &       \\
    &   2   &   9  &       &   8  \\\hline

\end{tabular}\]
Here, we have
$$\Delta_x F(1|1, 1) = P(Y=1|X=2, W=1 ) - P(Y=1|X=1, W=1) = 0.208;~~\mbox{and}$$
$$\Delta_x F(1|1, 2) =  P(Y=1|X=2, W=2 ) - P(Y=1|X=1, W=2) =-0.1.$$
That is, the distribution dependence is not homogeneous. Also,
 from the marginal table of $Y$ and $X$,
\[\Delta_x F(1|1) = P(Y=1|X=2) - P(Y=1|X=1) = 0.068 \neq \Delta_xF(1|1,w),\]
so that the distribution dependence function is not collapsible
over $W$. However, from the marginal table of $X$ and $W$,
\[\begin{tabular}{|cc|lcr|}
\hline
       &        &       &   W   &   \\
     &      &   1   &           &   2   \\
\hline
       &    1   &   12 &           & 10    \\
  X  &      &       &            &      \\
       &    2   &   24  &          &  20  \\
\hline
\end{tabular}\]
it can be seen that $X\perp W$ and
\begin{eqnarray*}
E_{W|X=1}\left( \Delta_{x}F(1|1, W) \right)
%&=&\displaystyle\sum_{w}\dfrac{\partial F(y|x, w)}{\partial x} f_{W|X}(w|x)\\
&=&\displaystyle\sum_{w}(\Delta_{x} F(1|1, w))f_{W|X}(w|x) \\
&=& \Delta_{x}F(1|1,1)f_W(1) + \Delta_{x}F(1|1,2)f_W(2) = 0.068\\
&=& \Delta_{x}F(1|1).
\end{eqnarray*}
Therefore, the distribution dependence function is A-collapsible
with respect to the background variable $W$.}
\end{example}

{\definition The distribution dependence function $\dfrac{\partial
F(y|x, w)}{\partial x}$ is A-collapsible over  W if
\begin{eqnarray} \label{eq3.1}
E_{W\mid X=x}\left(\dfrac{\partial F(y|x,W)}{\partial x}\right) = \dfrac{\partial F(y|x)}{\partial x},~\mbox
{for all y and x}.
\end{eqnarray}}
\noindent The above definition is a natural extension of simple
collapsibility of distribution dependence. In fact, when
$\dfrac{\partial F(y|x,w)}{\partial x}$ is homogeneous over $W$,
A-collapsibility reduces to collapsibility. Note also that
\eqref{eq3.1} is equivalent to having the second term on the
right-hand side of \eqref{eq2.2} zero.

\noindent The next result shows that the conditions of Cox \&
Wermuth (2003) are sufficient for A-collapsibility.

{\theorem \label{thm3.1} (a): Either of the conditions\\
(i) $Y\perp W\mid X$ or\\
(ii) $W\perp X$\\
is sufficient for the distribution dependence function
$\dfrac{\partial F(y|x, w)}{\partial x}$ to be A-collapsible
 over  the background variable $W$.

\noindent (b): Conversely, if $W$ is binary, say $W\in \{1,2\}$,
then the condition (i) or (ii) is
 also necessary.}

\begin{remark} {\em \label{con-rem} As pointed out by Cox \& Wermuth (2003, p.~940) and Xie \emph{et al.~}(2008, p.~1174),
the conditions  (i) and (ii) of collapsibility, and in general
A-collapsibility, are useful for the data analysis (e.g.,
contingency table), causal inference, observational studies and
the design of experiments. For example, the condition  (ii) may be
ensured by the proportional allocation of individuals to
treatments, though condition (i) involving the response can not be
ensured, during the planning stage of a study. However, one may
use a statistical test based on the full data, to check if
condition (i) is satisfied. }
\end{remark}

\noindent An example follows showing that the claim  in  Part (b)
of Theorem \ref{thm3.1} is in general not valid.

\begin{example} \label{excon}{\em
Let $W \in \{1, 2, 3 \}$ and $ 0< X < 2.$ In this case,
\eqref{veq3.2} reduces to
\begin{eqnarray} \label{neq3.1}
\bigg(F(y|x,1)- F(y|x, 3)\bigg) \frac{\partial f(1|x)}{\partial x}
+ \bigg(F(y|x,2)- F(y|x, 3)\bigg) \frac{\partial f(2|x)}{\partial
x} =0, ~~ \mbox{for  all}~ y~ \mbox{and}~ x.
\end{eqnarray}

Consider now the conditional distributions defined by
\begin{eqnarray} \label{neq3.2}
f(w|x)= \left\{
\begin{array}{lll}
 \displaystyle{ \frac{1+x}{8}}, & \mbox{for} & w=1, \\
  \displaystyle{ \frac{2-x}{4}}, & \mbox{for} & w=2, \\
  \displaystyle{ \frac{3+x}{8}}, & \mbox{for} & w=3 ,
\end{array}
\right.
\end{eqnarray}
where  $0<x<2.$  Then
\begin{equation}
\frac{\partial f(1|x)}{\partial x}= \frac{1}{8}; ~~ \frac{\partial
f(2|x)}{\partial x}= \frac{-1}{4}.
\end{equation}

\noindent Assume that $(Y|x,w) \sim U(w, w+x)$ so that
\begin{equation}
 F(y|x, w) = \frac{y-w}{x}, ~ w< y <w+x,
\end{equation}
where $x \in (0, 2)$ and  $w \in \{1, 2, 3 \}.$

\noindent The above conditional distributions  $F(y|x, w)$ and
$f(w|x)$ satisfy \eqref{neq3.1}, but neither $Y \perp W|X$ nor $X
\perp W$ is satisfied. }
\end{example}

\noindent Next, we construct, as asked by the reviewers, an
example where $Y$ has common support with respect to different
values of $X$ and $W$ and yet demonstrates the phenomenon of
A-collapsibility. Henceforth, $\phi(z)$ and $\Phi(z)$ respectively
denote the density and the distribution of $Z \sim N(0, 1).$

\begin{example} \label{ex-lr} {\em
Consider the linear regression model
\begin{equation}\label{eq2.4n}
 Y= \alpha_1X +\alpha_2 W + \alpha_3 XW + \epsilon,
\end{equation}
where $\epsilon \perp (X, W)$ and $\epsilon \sim N(0, \sigma^2)$.\\
Then
    $$ (Y|x, w) \sim N(m(x,w), \sigma^2),$$
where $ m(x, w)=\alpha_1x +\alpha_2 w + \alpha_3 xw$.

\noindent Therefore,
\begin{equation}\label{eq2.5n}
 \dfrac{\partial F(y|x, w)}{\partial x}=\Big(\frac{-1}{\sigma}\Big)(\alpha_1+ \alpha_3w)
 \phi \Big(\dfrac{y-m(x,w)}{\sigma}\Big).
\end{equation}

\noindent Assume now $W\perp X$ and $W \sim N(0, 1)$. Also, let  $
v^{2}(x, \sigma)= (\alpha_2 +\alpha_3x)^2+ \sigma^2.$ Then,
\begin{eqnarray} \label{eq2.6n}
E_{W\mid X=x}\left(\dfrac{\partial F(y|x,W)}{\partial x}\right)
&=&
 \int_{-\infty}^{\infty}\dfrac{\partial F(y|x,w)}{\partial x}
 f(w)dw \nonumber \\
 &=&\int_{-\infty}^{\infty} \Big(\dfrac{-1}{\sigma}\Big)(\alpha_1 + \alpha_3w)
 \phi\bigg(\dfrac{y- m(x, w)}{\sigma}\bigg) \phi (w)dw  \nonumber \\
 &=& \Big(\dfrac{-1}{v(x, \sigma)}\Big) \phi\bigg(\dfrac{y- \alpha_1x}{v(x,
 \sigma)}\bigg) \bigg[\alpha_1 + \dfrac{\alpha_3 (y-\alpha_1x)(\alpha_2+\alpha_3x)}{v^2(x, \sigma)}
 \bigg],
\end{eqnarray}
which follows using the results
\begin{eqnarray*} \label{eq2.7n}
 \int_{-\infty}^{\infty} \phi(a+bz) \phi(z) dz &=& s \phi(as); \\
\int_{-\infty}^{\infty} z\phi(a+bz) \phi(z) dz &=& ms \phi(as),
\end{eqnarray*}
where $ s= 1/ \sqrt{(1+b^2)}$ and $m= -ab s^2.$

\noindent On the other hand, from the model \eqref{eq2.4n} and the
assumption $W \sim N(0, 1)$, we have
\begin{eqnarray*} \label{eq2.8n}
  E( Y|x) &=& E_{W|x}(E(Y|x, W))= \alpha_1 x; \\
  V(Y|x) &=& E_{W|x}(V(Y|x, W)) + V_{W|x}(E(Y|x, W))\\
         &=& (\alpha_2 +\alpha_3x)^2+ \sigma^2 \\
         &=& v^2(x, \sigma).
\end{eqnarray*}
 Then  $(Y|x) \sim N(\alpha_1 x, v^{2}(x, \sigma))$ and it can be
seen that $\dfrac{\partial F(y|x)}{\partial x} $ equals the
right-hand side of \eqref{eq2.6n} and hence the A-collapsibility
holds. }
\end{example}

\noindent Note also from  \eqref{eq2.2} that A-collapsibility
holds if and only if
\begin{eqnarray}\label{eqs1}
\int F(y\mid x,w) \frac{\partial{f(w\mid x)}}{\partial x} \,dw =
0~~ \mbox{for all $(y,x)$}.
\end{eqnarray}

The following counter-example, which is the simplest one that we
have been able to find, shows that A-collapsibility can hold even
when neither condition (i) nor condition (ii) of Theorem
\ref{thm3.1} holds. Hence, these conditions are not necessary,
unless the background variable $W$ is binary.

\begin{example} \label{ex3} {\em
Let $Y$, given $X=x$ and $W=w$, follow uniform $U(0,
(x^2+(w-x)^2)^{-1})$ so that
\begin{eqnarray}
F(y|x,w) = y(x^2+(w-x)^2), ~~0<y<(x^2+(w-x)^2)^{-1}.
\end{eqnarray}
Assume also $(W|X=x)\sim N(x,1)$ so that
\begin{eqnarray}
\frac{\partial}{\partial x}f(w|x) = - \phi'(w-x) = (w-x)\phi(w-x).
\end{eqnarray}
% where $\phi(z)$ denotes the density of $N(0, 1)$ distribution.
 Hence,
\begin{eqnarray}
\int F(y|x,w)\frac{\partial}{\partial x} f(w|x)dw
&=& y\int_{-\infty}^{\infty}(x^2+(w-x)^2)(w-x)\phi(w-x)dw\nonumber\\
&=& y\bigg[x^2\int_{-\infty}^{\infty}(w-x)\phi(w-x) dw + \int_{-\infty}^{\infty} (w-x)^3\phi(w-x)dw\bigg]\nonumber\\
&=& y\bigg[x^2\int_{-\infty}^{\infty} t\phi (t) dt + \int_{-\infty}^{\infty} t^3\phi(t)dt\bigg]\nonumber\\
&=& 0, ~~\mbox{for all $(y,x).$}
\end{eqnarray}
Thus, from (\ref{eqs1}), A-collapsibility  over $W$ holds, but
neither condition (i) nor condition (ii) is satisfied.}
\end{example}

The following result provides a necessary condition for A-collapsibility.  It shows also that the A- collapsibility
 of distribution dependence implies the  A-collapsibility of density dependence.

\begin{propos} \label{thm3.2}
Suppose $F(y|x,w)$ and $F(y|x)$ admit continuous
mixed partial derivatives (with respect to $y$ and $x$). Then a
necessary condition for A-collapsibility of the distribution dependence function over
$W$ is
\begin{eqnarray}
 E_{W|X=x}\Bigg(\frac{\partial f(y|x,W)}{\partial x}\Bigg)= \frac{\partial f(y|x)}
  {\partial x},~~ \forall~ (y,x). \label{veqn3.1}
\end{eqnarray}
\end{propos}

For instance, the A-collapsibility of density dependence also
holds in Example \ref{ex3}.

\vone \noindent {\bf 4. Average collapsibility of quantile regression coefficients}

\vone \noindent For brevity, we assume in this section that all the random
variables under consideration are continuous with finite
variances. Consider the conditional (linear) regression model,
namely,
\begin{eqnarray}\label{Coch2}
E(Y|X=x,W=w) = \alpha_2 + \beta_{yx.w}x+ \beta_{yw.x}w.
\end{eqnarray}
Assume the marginal model is also linear and is defined by
\begin{eqnarray}\label{Coch1}
E(Y|X=x) = \alpha_1 + \beta_{yx}x.
\end{eqnarray}
Cochran (1938) proved the following relation for marginal and
conditional regression coefficients:
\begin{eqnarray}\label{Cochran}
\beta_{yx} = \beta_{yx.w} + \beta_{yw.x}\beta_{wx},
\end{eqnarray}
where $\beta_{yx}$ denotes the linear regression coefficient of
$Y$ on $X$, and $\beta_{yx.w}$ denotes corresponding coefficient
of $Y$ on $X$, when
 $W=w$ is fixed, and so forth.  Equation (\ref{Cochran}) decomposes the effect of a unit change in $X$ on the
 response variable $Y$ into two parts, the first being the effect with $W$ fixed, and the second a product
  of two effects: the effect of a unit change in $X$ on the moderating variable $W$, times the effect of a unit
  change in $W$ on the response $Y$ when $X$ is fixed.  Cox (2007) noted that (\ref{Cochran}) is essentially
  the formula for the total derivative of $y=y(x, w(x))$, namely,
 \[\frac{dy}{dx}=\frac{\partial y}{\partial x}+\frac{\partial y}{\partial w}\frac{dw}{dx}\]
 and hence could be extended to the  more general setting of quantile regression coefficients, which we now describe.
  Given $0<\eta<1$,
the function $y_{\eta} = y_{\eta}(x)$ satisfying $F(y_{\eta}|x) =
\eta$ is called $\eta$-th quantile function. The function
\begin{eqnarray}
q_x(y|x) = \displaystyle\frac{-\frac{\partial }{\partial x}F(y|x)}{f(y|x)}
\end{eqnarray}
is called the quantile regression coefficient (equation (2) of Cox, 2007).  Note that
\[\frac{\partial}{\partial x}y_\eta(x)=q_x(y_\eta(x)|x)\]
by implicit differentiation.  Hence, the quantile regression
function describes the effect of a unit change
 in $X$ on quantiles of $Y$.
Similarly,
\begin{eqnarray}
q_x(y|x,w) = \displaystyle\frac{-\frac{\partial }{\partial
x}F(y|x,w)}{f(y|x,w)}
\end{eqnarray}
represents the conditional quantile regression coefficient. Cox
(2007, p.757) established that
\begin{equation}\label{q1}
q_x(y|x) = E_{W|y,x}\{\delta(y|x,W)\} ,
\end{equation}
where $\delta(y|x,w)=q_x(y|x,w) + q_w(y|x,w)q_x(w|x)$ represents the total effect on quantiles of $Y$ of a unit
change in $X$, calculated at $(x,w)$.
When $\delta(y|x,w)$ does not depend on $w$, Cox (2007) noted that
\begin{eqnarray}\label{eq3.9}
q_x(y|x) = \delta(y|x,w),
\end{eqnarray}
a result similar to that of Cochran (1938).  Our interest lies in the quantile regression coefficients $q_x(y|x)$
and $q_x(y|x,w)$.

{\definition The quantile regression coefficient $q_x(y|x, w)$ is A-collapsible over $W$ if
\begin{eqnarray}
q_x(y|x) = E_{W|y,x}(q_x(y|x,W)).
\label{eq4.0}
\end{eqnarray}}
The next result shows that  conditions (i) and (ii) of Cox \& Wermuth (2003) are sufficient for A-collapsibility.

{\theorem \label{thm4.1} The quantile regression coefficient $q_x(y|x, w)$ is A-collapsible over
$W$ if (i) $ Y\perp W|X$ or (ii) $ W\perp X.$}\\

\noindent {\bf Example \ref{ex-lr}} (continued). Consider Example
\ref{ex-lr} discussed earlier, where
\begin{eqnarray}
 F(y|x, w) = \Phi \Big( \frac{y-m}{\sigma}\Big) \label{neweqn4.1}
\end{eqnarray}
and $X>0$ is independent of $W \sim N(0, 1).$ By Theorem
\ref{thm4.1}, A-collapsibility of $q_{x}(y|x, w)$ holds.

\noindent Let, as before, $v^2(x)= (\alpha_2 + \alpha_3x)^2 +
\sigma^2.$ It can be seen that in this example,
\begin{eqnarray}
 F(y|x) = \Phi\Big(\frac{y-\alpha_1}{v(x)}\Big) \label{neweqn4.2}
\end{eqnarray}
and that the conditional density  of $W$ given $Y$ and $X$ is
\begin{eqnarray}
f(w|y,x)&=& \frac{f(y|w,x)f(w|x)}{f(y|x) }\nonumber \\
  &=& \frac{1}{s} \phi\Big( \frac{w-\eta}{s}\Big), \label{neweqn4.3}
\end{eqnarray}
where $s= \sigma /v,$ and $ \eta = (y- \alpha_1 x)(\alpha_2 +
\alpha_3 x) /v^2(x).$ Thus, $f(w|y,x)$ belongs to a two-dimensional
regular exponential family (Johansen, 1979).

\vone We next show, in general, that the converse of Theorem
\ref{thm4.1} is not true.
 Also, let $S_{yx}$ denote the support of $(Y,X)$. Note from
(\ref{eq4.2}), A-collapsibility holds
\begin{eqnarray}\label{eqn410}
 &\Longleftrightarrow & \int q_w(y|x,w)q_x(w|x)dF(w|y,x) = 0, ~\forall~(y, x)\in
 S_{yx}\\
 &\Longleftrightarrow&
\displaystyle\int(q_w(y|x,w)q_x(w|x))f(y|x,w)\frac{f(w|x)}{f(y|x)}dw = 0 \nonumber  \\
&\Longleftrightarrow & \int \frac{\partial}{\partial
w}F(y|x,w)\frac{\partial}{\partial x}F(w|x)dw = 0,~ \forall~(y,
x)\in S_{yx}.
 \label{eqn411}
\end{eqnarray}
The above fact is used to construct the following counter-example.

\example\em{Let $X>0$ and $W$ be real-valued continuous random
variables with
\begin{eqnarray*}
F(w|x) = \Phi\Big(\frac{w}{x}\Big),~ x>0,~ w\in \mathbb{R},
\end{eqnarray*}
so that
\begin{eqnarray*}
\frac{\partial}{\partial x}F(w|x) =
-\frac{w}{x^2}\phi\Big(\frac{w}{x}\Big).
\end{eqnarray*}
%where $\Phi$ and $\phi$ denote respectively the distribution  and the density function of $Z\sim N(0,1)$.\\
Also, let
\begin{eqnarray*}
F(y|x,w) = \frac{y+x-w}{2x},~ w-x<y<w+x,
\end{eqnarray*}
so that $Y$, given $X=x$ and $W=w$, follows uniform $U(w-x,~ w+x)$ and
\begin{eqnarray*}
\frac{\partial}{\partial w}F(y|x,w) = -\frac{1}{2x}, ~w-x<y<w+x.
\end{eqnarray*}
Then
\begin{eqnarray*}
\int_{-\infty}^{\infty} \frac{\partial}{\partial w}F(y|x,w) \frac{\partial}{\partial x}F(w|x)dw &=
&\frac{1}{2x^2}\int_{-\infty}^{\infty}\frac{w}{x}\phi\Big(\frac{w}{x}\Big)dw\\
&=&\frac{1}{2x}\int_{-\infty}^{\infty}t\phi(t)dt\\
&=& 0, ~\mbox{for all} ~(y,x)\in S_{yx}.
\end{eqnarray*}}
Using (\ref{eqn411}), A-collapsibility over $W$ holds. But,
neither condition (i) nor condition (ii) is satisfied. \vone

\noindent Next, we identify a class of conditional distributions
 of $W$ given $(Y, X)$, in view of (\ref{eqn410}), for which condition (i) or condition (ii) is also necessary.

\begin{theorem} \label{thm4.2} Let $W>0, \theta= \theta(y, x)$ and $(W|y,x)$ have density of the form
\begin{eqnarray}
f(w|y, x) = \frac{1}{\lambda(\theta)} e^{-\theta w} \nu(w),
\label{eqn5.1}
\end{eqnarray}
for some $ \lambda(\theta) > 0$, $ \nu(w) > 0$ and $ (y, x) \in
S_{xy}$.
 Then condition (i) or (ii) of
Theorem \ref{thm4.1} is also necessary.
\end{theorem}

\noindent Observe that the density $ f(w|y,x)= \lambda e^{-\lambda w}, ~ w>
0$, for some $\lambda = \lambda(y, x)>0$ and for all $ (y, x) \in
S_{xy}$,  is of the form given  in \eqref{eqn5.1}.

\noindent As another example, consider the binomial distributions
with ($0 \leq w \leq x $)
\begin{eqnarray}\label{eqn5.2}
f(w|y, x) &=& {x \choose w} y^{w} (1-y)^{x-w}  \nonumber \\
          &=& \frac{ {x \choose w} e^{-\theta w}}
          {(1+e^{-\theta})^x},
\end{eqnarray}
for $x \in \{1,2, \cdots \}$, $ y\in(0, 1)$ and  $ \theta = - ln
(y/(1-y)).$ This family of distributions is also of the form in \eqref{eqn5.1}. \\

\noindent Finally,  we briefly address the multivariate case.  As
discussed in Cox \& Wermuth (2003) and Xie \emph{et al.~}(2008),
the multivariate response $Y$ may be considered
 by treating one component at a time and similarly the multivariate $X$ may also be considered
 one contrast at a time, while keeping other components fixed.
 Therefore, as suggested by a referee, we consider here only the case where the covariate $W $
 is a random vector.

Let $W = (W_{1}, W_{2})$, where $W_{1}$ has $q~ (< p)$ components
and $W_{2}$ has ($p-q$) components. The definition of
A-collapsibility of  a measure of association remains the same,
except that $W$ is now a $p$-variate random vector. We now have
the following result.

\begin{theorem} \label{thm4.4} Let $W_{1} \perp W_{2}|X$ . Then
the distribution dependence function ${\partial F(y|x,
w)}/{\partial x}$  and the quantile regression coefficient
$q_x(y|x, w)$ are A-collapsible over $W$ if $(i)\hspace{1mm} Y
\perp W_{1}|(X, W_{2})$ and
 $(ii)\hspace{1mm} X \perp W_{2}$ hold.
\end{theorem}

\noindent When  the distribution dependence function
$\dfrac{\partial F(y|x, w)}{\partial x}$is homogeneous over $w_2$,
Xie \emph{et al.~}(2008, Theorem 5) proved its uniform
collapsibility.

\vone
\noindent {\bf Acknowledgements}

\noindent The author expresses his deep gratitude to Professor Mark M.
Meerschaert for several helpful discussions and
encouragements. The author is  grateful to the associate editor
for his extensive and detailed report with useful suggestions and
also to the referees for their critical comments, which have led
to significant improvements.
%A part of this work was
%done while the  author was visiting the Department of Statistics
%and Probability, Michigan State University, USA.
This research is partially
supported by a DST research grant No. SR/MS:706/10.

\newpage
\noindent {\bf  References}

\vone \noindent
Apostol, T. M. (1962). {\it Calculus.} Vol. II.  Blaisdell Publishing Company, New York.\\
Bishop, Y. M. M. (1971). Effects of collapsing multidimensional contingency tables. {\it Biometrics}, {\bf 27}, 545-562.\\
Cochran, W. G. (1938). The omission or addition of an independent
variable in multiple linear regression.
{\it J. R. Statist. Soc. Suppl.}, {\bf 5}, 171-176.\\
Cox, D. R. (2003). Conditional and marginal association for binary random variables.
{\it Biometrika}, {\bf 90}, 982-984.\\
Cox, D. R. (2007). On a generalization of a result of W. G. Cochran.
{\it Biometrika}, {\bf 94}, 755-759.\\
Cox, D. R. \& Wermuth, N. (2003). A general condition for avoiding effect reversal after marginalization.
{\it J. R. Statist. Soc. B}, {\bf 65}, 937-941.\\
%Ducharme, G. R. and Lepage, Y. (1986). Testing collapsibility in contingency tables.
%{\it J. R. Statist. Soc. B}, {\bf 48}, 197-205. \\
Geng, Z. (1992). Collapsibility of relative risk in contingency tables with a response variable.
{\it J. R. Statist. Soc. B}, {\bf 54}, 585-593.\\
Johansen, S. (1979). {\it Introduction to the theory of regular
exponential  families.}  Lecture Notes {\bf 3}, Institute of
Mathematical Statistics, University of Copenhagen.\\
Koenker, R. (2005). {\it Quantile regression}. Cambridge University Press, Cambridge.\\
Ma, Z., Xie, X. \& Geng, Z. (2006). Collapsibility of distribution dependence.
{\it J. R. Statist. Soc. B}, {\bf 68}, 127-133.\\
Mantel, N. \& Haenszel, W. (1959). Statistical aspects of the analysis of data from retrospective studies of disease.
{\it J. Natn. Cancer Inst.}, {\bf 22}, 719–748.\\
Vellaisamy, P. \&  Vijay, V. (2007). Some collapsibility results for n-dimensional contingency tables.
{\it Ann. Inst. Statist. Math.}, {\bf 59}, 557-576.\\
Vellaisamy, P. \&  Vijay, V. (2008). Collapsibility of regression coefficients and its extensions.
{\it J. Statist. Plann. Inference}, {\bf 138}, 982-994.\\
Vellaisamy, P. \&  Vijay, V. (2010). Collapsibility of contingency
tables based on conditional models.
{\it J. Statist. Plann. Inference}, {\bf 140}, 1243-1255.\\
Wermuth, N. (1987). Parametric collapsibility and the lack of moderating effects in contingency tables
with a dichotomous response variable. {\it J. R. Statist. Soc. B}, {\bf 49}, 353-364.\\
Wermuth, N. (1989). Moderating effects of subgroups in linear
models.
{\it Biometrika}, {\bf 76}, 81-92.\\
Whittemore, A. S. (1978). Collapsibility of multidimensional contingency tables.
{\it J. R. Statist. Soc. B}, {\bf{40}}, 328-340.\\
Xie, X., Ma, Z. \& Geng, Z. (2008). Some association measures and
their collapsibility. {\it Statist. Sinica}, {\bf 18}, 1165-1183.

\vone
\noindent P. Vellaisamy, Department of Mathematics, Indian Institute of Technology Bombay,
Powai, Mumbai-400076, India.

\noindent Email: pv@math.iitb.ac.in

\vone

\noindent {\bf Appendix: Proofs}

\noindent {\it Proof of Theorem \ref{thm3.1}}.
%Let $Y$ and $X$ be continuous
First assume condition (i) holds. Then
\begin{eqnarray*}
E_{W|X=x}\left(\dfrac{\partial F(y|x,W)}{\partial
x}\right)=E_{W|X=x}\left(\dfrac{\partial F(y|x)} {\partial
x}\right)=\dfrac{\partial F(y|x)}{\partial x}
\end{eqnarray*}
and hence A-collapsibility holds.

\noindent Assume next condition (ii) holds. Then
\begin{eqnarray*}
\dfrac{\partial F(y|x)}{\partial x}&=&\dfrac{\partial }{\partial x}\left[\int F( y|x,w)dF_{W|X}(w|x)\right]\\
&=&\displaystyle\int \left(\frac{\partial}{\partial x}F(y|x, w)\right)dF_{W}(w)\\
&=&\displaystyle\int_{w} \Big(\frac{\partial F(y|x,w)}{\partial x}\Big) dF_{W|X}(w|x)\\
&=& E_{W|X=x}\left(\dfrac{\partial F(y|x,W)}{\partial x}\right),
\end{eqnarray*}
showing again that A-collapsibility holds.

As to the converse, let $W$ be discrete and $$E_{W\mid
X=x}\left(\dfrac{\partial F(y|x,W)}{\partial x}\right) =
\dfrac{\partial F(y|x)}{\partial x}$$ hold for all $y$ and $x$.
Then,
\begin{eqnarray}
\sum_w \Big(\dfrac{\partial F(y|x,w)}{\partial x}\Big)f_{W|X}(w|x)
&=&\frac{\partial}{\partial x} \Big\{ \sum_{w} F(y|x,w)f_{W|X}(w|x)\Big\}\nonumber\\
&=&\sum_{w}f_{W|X}(w|x)\frac{\partial}{\partial x} F(y|x,w) \nonumber\\
&&+ \sum_{w} F(y|x, w)\frac{\partial}{\partial x}f_{W|X}(w|x).
\label{case1}
\end{eqnarray}
Hence,
\begin{eqnarray}
\sum_{w} F(y|x, w) \dfrac{\partial}{\partial x}f_{W|X}(w|x)= 0,
\:\; {\rm for \ all} \; x, y. \label{veq3.2}
\end{eqnarray}
Since $w \in \{1, 2 \}$ is binary, we have
\begin{eqnarray*}
 \frac{\partial}{\partial x} f_{W|X}(2|x)=
-\frac{\partial}{\partial x} f_{W|X}(1|x)
\end{eqnarray*}
and hence we get from (\ref{veq3.2}),
\begin{eqnarray*}
\{F(y|x,1)-F(y|x,2)\}\frac{\partial}{\partial x}f_{W|X}(1|x)=0,
~\mbox{for all $y$ and $x$.}
\end{eqnarray*}
Thus, we get
 $ F(y|x,1)= F(y|x,2)$ or $\dfrac{\partial}{\partial x}f_{W|X}(1|x)=0$, which are equivalent to \\
$$Y\perp W\mid X\quad \mbox{or}\quad X\perp W,$$
respectively.
%To see that $\frac{\partial}{\partial x}f_{W|X}(1|x)=0$ implies
%$X\perp W$, note that if $P(W=2|X=x)=P(W=1|X=x)$ for all $x$, then
%$P(W=2,X=x)=P(W=1,X=x)=0.5P(X=x)$ for all $x$, and $X\perp W$
%follows easily.

\noindent {\it Proof of Proposition 1.} We give the proof for the
case of discrete $W$. Assume A-collapsibility holds. Then from
(\ref{veq3.2}),
\begin{eqnarray}
\sum_{w} F(y|x, w) \dfrac{\partial}{\partial x}f_{W|X}(w|x)= 0,
\:\; {\rm for \ all} \; x, y. \label{veqn3.2}
\end{eqnarray}

\noindent Also,
\begin{eqnarray}
\sum_{w} F(y|x,w) f(w|x)  =  F(y|x), \ \forall \ (y,x).
\label{veqn3.3}
\end{eqnarray}

\noindent Differentiating (\ref{veqn3.3}) with respect to $x$,
using  (\ref{veqn3.2}), and then differentiating with respect to
$y$, we get

%\begin{eqnarray}
%\sum_{w} \frac{\partial F(y|x,w)}{\partial x} f(w|x)  =
%\frac{\partial F(y|x)}{\partial x}, \ \forall \ (y,x).
%\label{veqn3.4}
%\end{eqnarray}

%\noindent Differentiating (\ref{veqn3.4}) now with respect to
%$y$, we get
\begin{eqnarray}
\sum_{w} \frac{{\partial}^2 F(y|x,w)}{{\partial y \partial x}}
f(w|x) =  \frac{{\partial}^2 F(y|x)}{{\partial y \partial} x}, \
\forall \ (y,x). \label{veqn3.5}
\end{eqnarray}

\noindent Since  $F(y|x)$ has continuous mixed partial derivatives
(Apostol, 1962, p. 214), we have

%\[\frac{\partial^2}{\partial y\partial x}F(y|x)=\frac{\partial^2}{\partial x\partial y}F(y|x)
%=\frac{\partial}{\partial x}f(y|x)\] (e.g., see Apostal (1962) p.
%214), and hence
$$ \frac{{\partial}^2 F(y|x)}{\partial y \partial x} = \frac{\partial f(y|x)}{\partial x};
 \ \ \  \frac{{\partial}^2 F(y|x,w)}{\partial y \partial x} = \frac{\partial f(y|x, w)}{\partial x}, \ \ \forall \ (y,x). $$
Substituting the above facts in (\ref{veqn3.5}), we obtain
\begin{eqnarray}
\sum_{w} \frac{\partial f(y|x,w)}{\partial x} f(w|x)
=\frac{\partial f(y|x)}{\partial x}, \ \forall \ (y,x),
\label{veqn3.6}
\end{eqnarray}
which proves the result.

\noindent{\it Proof of Theorem \ref{thm4.1}}.  From Cox's result
(\ref{q1}),
\begin{eqnarray}
\label{eq4.1} q_x(y|x)&=&E_{W|y,x}(q_x(y|x,W))\\\nonumber
&\Longleftrightarrow&  E_{W|y,x}(q_w(y|x,W)q_x(W|x)) = 0
\nonumber\\
&\Longleftrightarrow&  \int(q_w(y|x,w)q_x(w|x))dF(w|y,x) = 0,
~\mbox{for all }~(y, x).
 \label{eq4.2}
\end{eqnarray}
If condition (i) holds, then since
\begin{eqnarray}
Y\perp W|X \Longleftrightarrow F(y|x,w)=F(y|x)~\mbox{for all
$y$, $x$ and $w$},
\end{eqnarray}
we have $q_w(y|x,w) = 0$. Hence, (\ref{eq4.1}) holds.\\
If condition (ii) $W\perp X$ holds, then,
\begin{eqnarray*}
F(w|x) &=& F(w) ~\mbox{for all} ~(w,x)\\
&\Rightarrow& q_x(w|x) = 0  ~\mbox{for all} ~(w,x),
\end{eqnarray*}
which in turn proves (\ref{eq4.1}). This proves the result.

\noindent {\it Proof of Theorem \ref{thm4.2}}. Let
A-collapsibility of $q_{x}(y|x,w)$ hold. Then from (\ref{eqn410}),
\begin{eqnarray*}
\int_{0}^{\infty} q_w(y|x,w)q_x(w|x)dF(w|y,x) = 0, ~\mbox{for
all}~~(y, x)\in S_{yx}
\end{eqnarray*}
which implies
\begin{eqnarray*}
\int_{0}^{\infty} q_w(y|x,w)q_x(w|x) \nu(w) e^{-\theta w} dw = 0,
~\mbox{for all}~~(y, x)\in S_{yx}.
\end{eqnarray*}
By the uniqueness of the Laplace transform, we now have
\begin{eqnarray}
 q_w(y|x,w)q_x(w|x) = 0, ~\mbox{for all}~(y, x)\in
S_{yx}
\end{eqnarray}
which is equivalent to
            $$q_w(y|x,w) = 0,  ~\mbox{or} ~  q_x(w|x) = 0.$$
That is, condition (i) or (ii) holds.

\noindent{\it Proof of Theorem \ref{thm4.4}}. Note that
\begin{eqnarray}
E_{W|x}\left(\frac{\partial}{\partial
x}F(y|x,W)\right)&=&\int_{w_{2}}\int_{w_{1}}\left(\frac{\partial}{\partial
x}F(y|x,w)\right)\,dF(w_{1},w_{2}|x)\nonumber \\
&=& \int_{w_{2}}\int_{w_{1}}\left(\frac{\partial}{\partial
x}F(y|x,w_{1},w_{2})\right)\,dF(w_{2}|x)\,dF(w_{1}|x),\hspace{4mm}(\because \hspace{2mm} W_{1}\perp W_{2}|X)\nonumber \\
&=& \int_{w_{2}}\int_{w_{1}}\frac{\partial}{\partial
x}F(y|x,w_{2})\,dF(w_{2}|x)\,dF(w_{1}|x),
\hspace{4mm}(\because \hspace{2mm} Y\perp W_{1}|(X,W_{2}))\nonumber \\
&=& \int_{w_{2}}\frac{\partial}{\partial x}F(y|x,w_{2})\,dF(w_{2}|x)\nonumber \\
&=& E_{W_{2}|x}\left(\frac{\partial}{\partial x}F(y|x,W_{2})\right)\nonumber \\
&=& \frac{\partial}{\partial x}E(Y|x) \hspace{4mm}\mbox{for all}
~x, \nonumber
\end{eqnarray}
by condition $(ii)$ and  Theorem \ref{thm3.1}. \\
The proof for the quantile regression coefficient $q_w(y|x,w)$
follows similarly and uses Theorem \ref{thm4.1}.

%\vtwo \noindent \today

\end{document}